\documentclass[a4paper,11pt]{article}
\usepackage{url}
\usepackage{color}
\usepackage{overpic}
\usepackage[colorinlistoftodos]{todonotes}
\usepackage{jinstpub} 
\usepackage{lineno}

\title{\boldmath Studying the GRAiNITA concept: first test beam results}

\author[a]{Sergey Barsuk,}
\author[d]{Oleg Bezshyyko,}
\author[a,b]{Ianina Boiaryntseva,}
\author[b]{Andrey Boyarintsev,}
\author[a]{Dominique Breton,}
\author[c]{Herv\'e Chanal,}
\author[b]{Alexander M. Dubovik,}
\author[d]{Larysa Golinka-Bezshyyko,}
\author[a]{Carlos Dominguez Goncalves,}
\author[c]{Yingrui Hou,}
\author[a]{Giulia Hull,}
\author[a,]{Miktat Imre,}
\author[a,d]{Denys Klekots,}
\author[a]{Jacques Lefran\c{c}ois,}
\author[a]{Jihane Maalmi,}
\author[c]{Magali Magne,}
\author[a]{Bernard Mathon,}
\author[c]{St\'ephane Monteil,}
\author[a]{Sebastien Olmo,}
\author[c]{David Picard,}
\author[a,1]{Marie-H\'el\`ene Schune,\note{Corresponding author.}}
\author[b]{Irina Tupitsyna}
\author[c]{and Mykhailo Yeresko.}
\affiliation[a]{Université Paris-Saclay, CNRS/IN2P3, IJCLab, Orsay, France}
\affiliation[b]{Institute for Scintillation Materials of the National Academy of Sciences of Ukraine, 60 Nauki Ave., Kharkiv 61072, Ukraine}
\affiliation[c]{Université Clermont-Auvergne, CNRS/IN2P3, LP-Clermont, 63177 Aubi\`ere, France}
\affiliation[d]{Kyiv National Taras Shevchenko University, 01033 Kyiv, Ukraine}

\emailAdd{marie-helene.schune@ijclab.in2p3.fr}

\abstract{Data collected over a two-day period in June 2024 at the CERN SPS H9 test beam using a small-scale GRAiNITA prototype have been analyzed to characterize the detector’s energy-resolution performance. 
The measurements allow 
for a first estimate of the constant term associated with detector non-uniformity. Although the evaluation is limited by the small prototype size and the use of pion beams, the results indicate that the non-uniformity-related constant term is significantly below 1\%. Furthermore, the test-beam data confirm that the contribution to the energy resolution arising from photo-electron statistics is approximately $1\%/\sqrt{E}$. These findings validate the expected calorimetric performance of the GRAiNITA concept and provide important input for the design and optimization of future full-scale detectors.}

\def\papercopyright{\the\year\ CERN for the benefit of the DRD Calo collaboration} 
\def\paperlicence{CC BY 4.0 licence}
\def\paperlicenceurl{https://creativecommons.org/licenses/by/4.0/}

\begin{document}

\maketitle
\flushbottom

\section{Introduction}

Access to the full physics potential of a future high-energy $e^+ e^-$ collider, such as the FCC-ee, particularly in the domain of flavour physics at the $Z^0$ pole, necessitates the development of innovative and cost-effective electromagnetic calorimeters. These detectors must ensure sufficient energy resolution to avoid limiting the reconstruction of decay modes involving photons, neutral pions, or electrons. 
For the innovative GRAiNITA calorimeter concept the successive layers of scintillator and absorber materials of
a conventional Shashlik detector are replaced by millimetric grains of high-atomic number and high-density
inorganic scintillator crystals, evenly distributed in a bath of transparent high-density liquid. The light yielded by the grains remains close to the production zone because of reflection/refraction at the grain liquid interface and can be collected towards the photodetectors by means of wavelength shifting
(WLS) fibres running along the detection length, with a design similar to a conventional sampling calorimeter. 

A small-scale GRAiNITA prototype, employing millimetric grains of ZnWO$_4$ crystals, was constructed to perform an initial validation of the detector concept. Indeed, ZnWO$_4$ is a well suited non-hygroscopic inorganic scintillator material for GRAiNITA of high density 7.87 g/cm$^3$), effective atomic number (61) with a good light yield ($\sim$ 9000 photoelectrons /MeV)~\cite{Grassman:1985JoL,Holl:12684} and grains can be obtained via spontaneous crystallization from a flux melt~\cite{bib:Yamada}.
Measurements using several hundred cosmic muons yielded approximately 400 photo-electrons for an estimated deposited energy of about 40 MeV~\cite{Barsuk:2023uva}, indicating the potential to achieve an energy resolution contribution from photo-electron statistics at the level of $1\%/\sqrt{E}$. Simulation studies with a simplified GEANT4 set-up~\cite{G4} indicate that the 1\% term degrades to approximately 2\%~\cite{bib:IEEE2021} as a result of event-to-event sampling fluctuations in the partition of deposited energy between the scintillator grains, which contribute to the detected signal, and the inter-grain heavy liquid, which does not.
This term is smaller than in usual sampling calorimeter due to the smallness of the grains and the high density of the scintillator. 

This publication has two goals: first to verify the stochastic term from larger and better controlled data samples and, more importantly, to perform a first investigation of the contribution to the energy resolution of the constant term due to potential non-uniformity response of the detector. Both aspects using the same small-scale GRAiNITA prototype were studied using muon and pion beams at CERN.

\section{Experimental setup}
\begin{figure}[htbp]
\centering
\includegraphics[width=.25\textwidth]{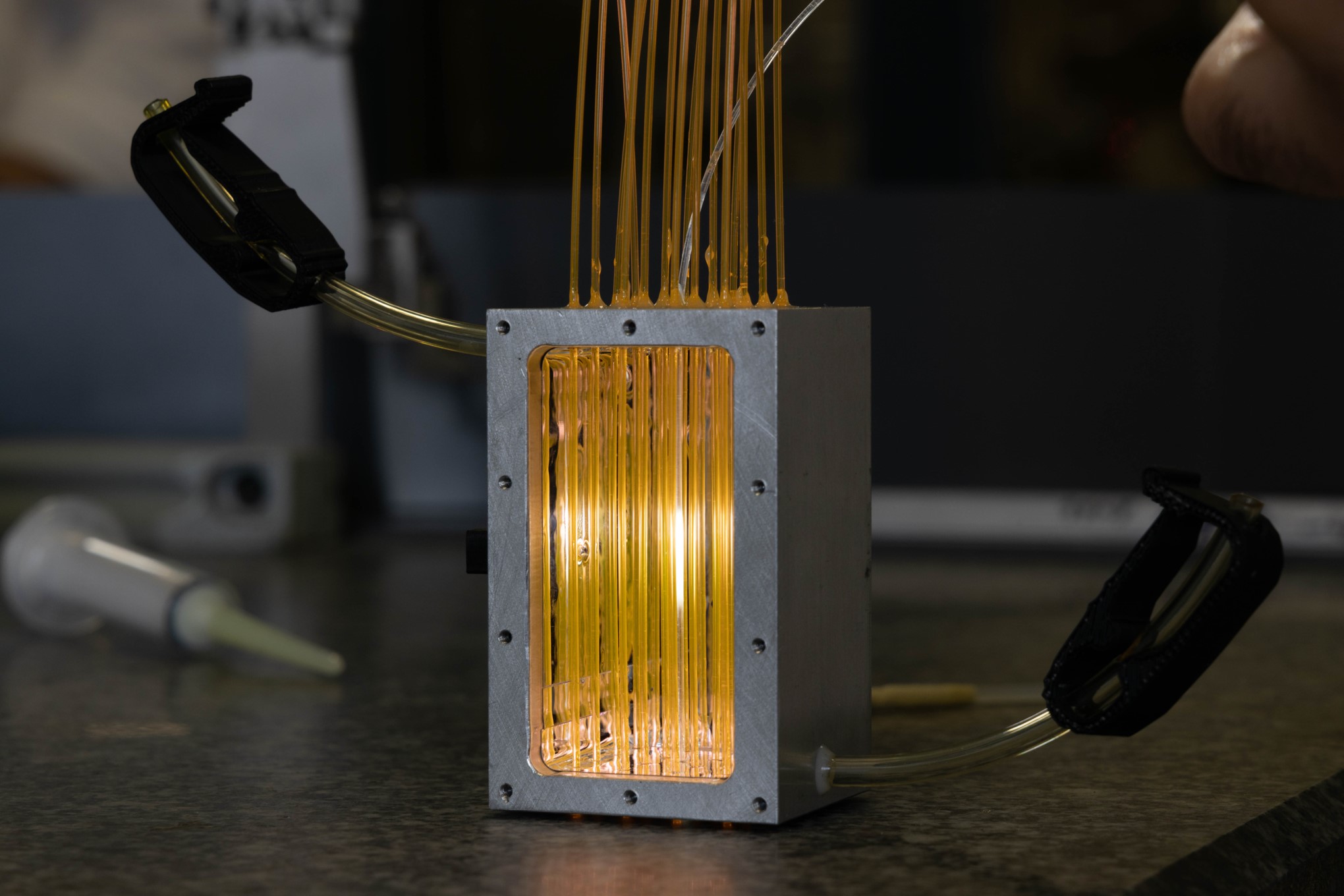}
\includegraphics[width=.15\textwidth]{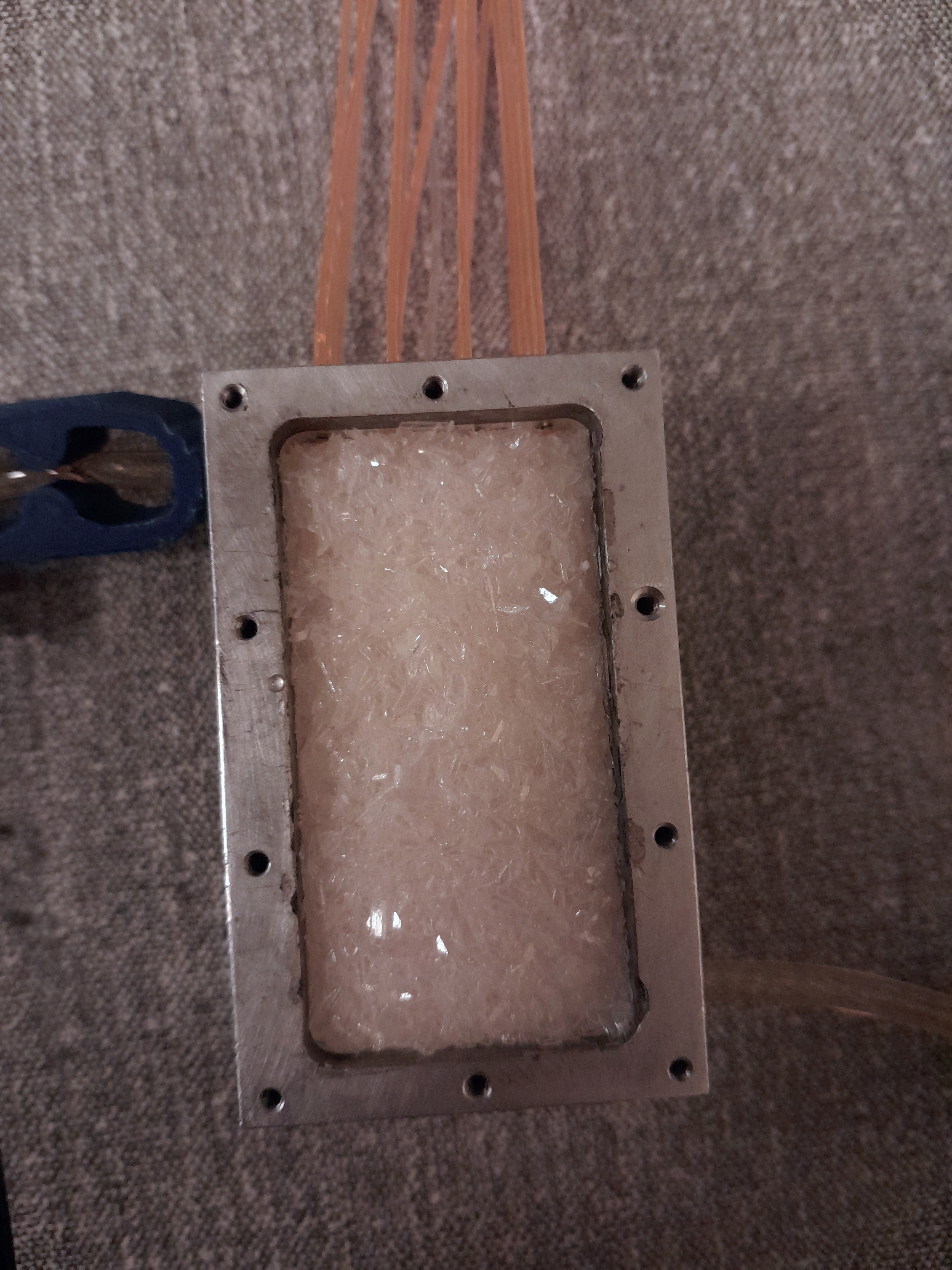}
 \includegraphics[width=0.5\textwidth]{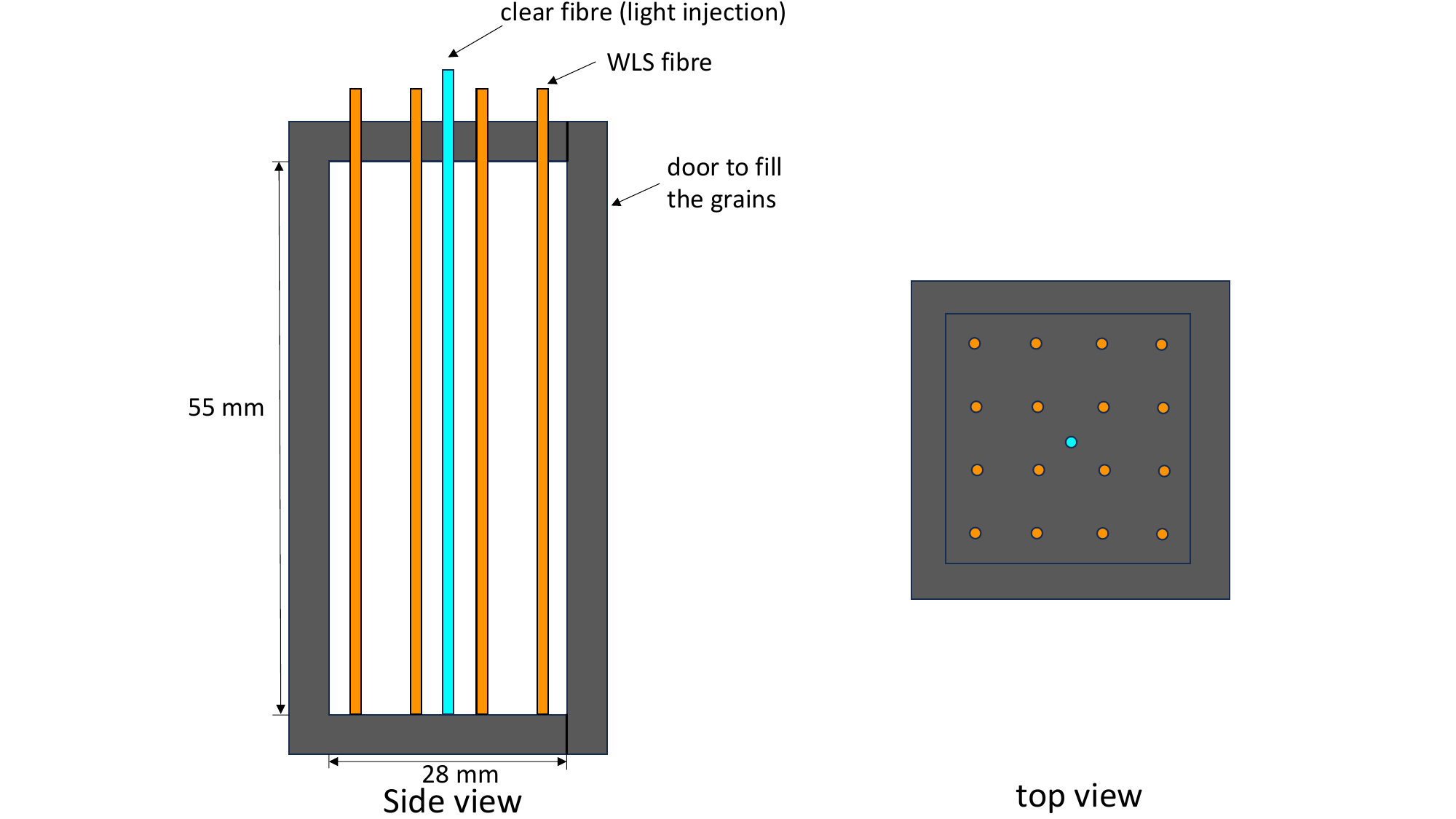}
\caption{The 16-channels GRAiNITA prototype (left: real, middle: real filled with grains, right: schematic).  \label{fig:theTroll}}
\end{figure}

The 16-channels GRAiNITA small-size prototype is shown in figure~\ref{fig:theTroll}. It consists of a metallic container with an active volume of $28 \times 28 \times 55$~mm$^3$ and it is equipped with 16 O-2(200) Kuraray WLS fibers (1 mm diameter)~\cite{bib:KurarayWLSOnline} arranged in square geometry and inter-spaced by 7 mm. In order to monitor the behaviour of the setup, a clear fiber, unpolished along 1 cm, is placed in the middle of the WLS ones and connected to a green LED\footnote{Broadcom HLMP-CM1A-560DD, wavelength from 500 to 540 nm.} to inject the light in the active volume. 
Each of the 16 WLS fibers of GRAiNITA is coupled to a 1.3 $\times$ 1.3 mm$^2$ SiPM S13360-1350E by Hamamastu ~\cite{bib:HamamastuOnline} mounted on a PCB for supply, gain stability and amplification. The signals from the SiPMs are digitized at 3.2 GigaSample/s with a WaveCatcher module~\cite{bib:WaveC}. The number of photo-electrons is recorded in a 25 $\mu$s windows corresponding to the scintillating decay time constant of the $\textrm{ZnWO}_4$ grains.

\begin{figure}[htbp]
\centering
 \includegraphics[width=0.55\textwidth]{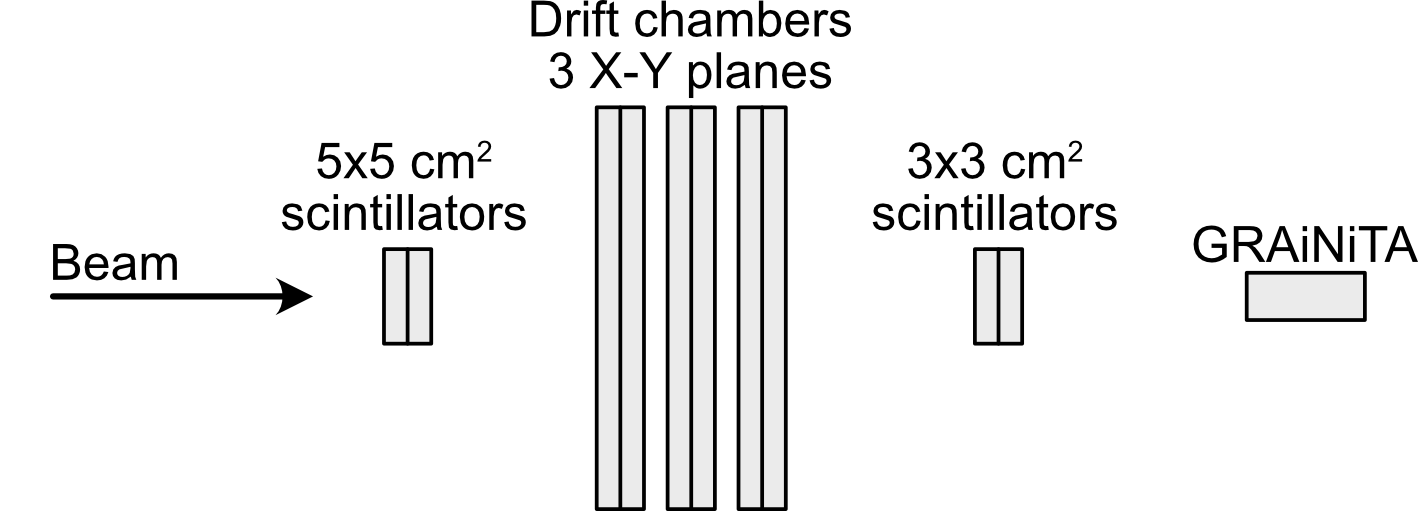}
 \hskip 1.3cm
 \includegraphics[width=0.35\textwidth]{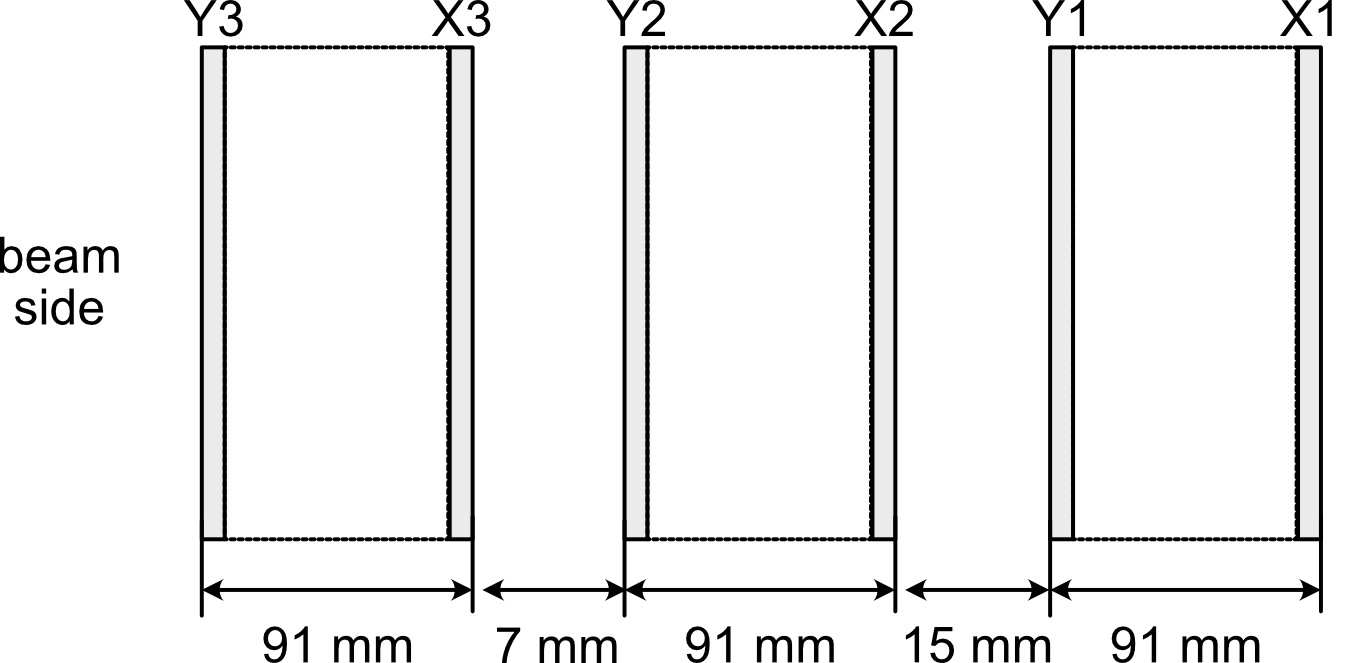}
\caption{Schematic view of the test beam setup (left) with a zoom on the DWC arrangement (right). \label{fig:setup}}
\end{figure}

Data-taking took place during two days at the SPS beam (H2 line) delivering muons and pion secondary beams. Due to unstable beam conditions, the energy was varying but remained above few GeV, enough to ensure stable conditions for the measurements described in the following. In total about 2 million of muons events and 48 millions of pions events were recorded. Given the smallness of the GRAiNITA prototype, the estimated interaction probability for the pions is approximately 20\%. 
The data acquisition was divided into two periods. In the first period (P1), ZnWO$_4$ grains were immersed in water, while in the second period (P2) they were immersed in a water-based sodium polytungstate solution, hereafter referred to as the heavy liquid, characterized by a refractive index of 1.5 and a density of $2.85 \ \mathrm{g . cm}^{-3}$. For both periods, data were collected using pion and muon beams. Toward the end of the acquisition, partial leaks of the heavy liquid were observed due to a container closure not completely liquid-tight. Consequently, we chose to focus on the data obtained with water to evaluate non-uniformity effects, while using the heavy-liquid data primarily for calibration purposes, since no significant differences in non-uniformity are expected between the two liquid configurations.
The experimental setup used is shown in figure~\ref{fig:setup} and consists of a set of 3 wire chambers (DWC) for reconstructing the beam tracks, 2 sets of dual scintillator for the triggering surrounding the GRAiNITA prototype.  

\section{Track reconstruction and checks \label{sec:hitmap}}
Given the reduced size of the GRAiNITA prototype, a simple track reconstruction assuming that the beam propagates perpendicular to the setup was performed. The resolutions were estimated comparing the measured $(x,y)$ coordinates for each of the three DWC and the predicted positions from the track reconstruction. A resolution of the order of 170 $\mu$m is achieved. 

During the whole data-taking process, random events without tracks in the DWC are recorded in order to estimate the dark noise. Its level is found to be of the order of ten counts per channel in the 25~$\mu$s time window and is subtracted for the analysis. 

\begin{figure}[ht!]
    \centering
    \includegraphics[width=0.6\textwidth]{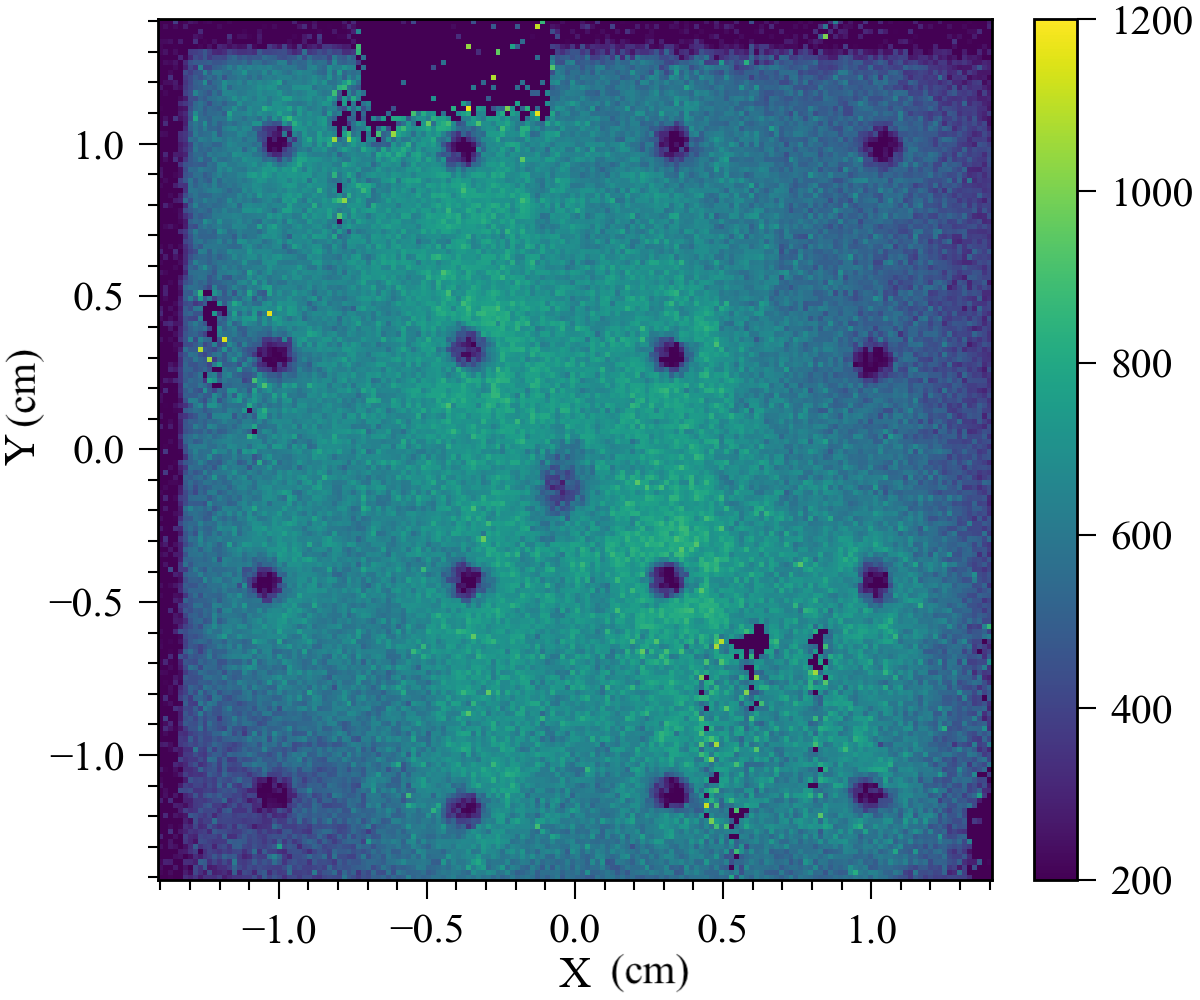}
    \caption{Mean number of photo-electrons per track for the sum of the 16 fibres.}
    \label{fig:calib:hitmap}
\end{figure}

A very first look at the detector performance can be done using the very large statistics pions runs recorded during P1. A plane located in the $z$ position corresponding to the centre of the DWCs is paved with bins of 170 $\mu$m width which is corresponding roughly to the resolution on the track position. For each bin, the mean value of the number of photo-electrons per track is computed for the sum of the 16 fibres. The obtained hitmap is shown in figure~\ref{fig:calib:hitmap}. The position of each of the 16 WLS fibres can clearly be seen as they do not produce scintillating light through the interaction with the beam particle. The clear fibre in the centre used to inject green light for monitoring purposes is more fuzzy. This is attributed to an internal bending of the fibre over the small prototype length. The last striking feature is a blind zone, in the upper region. This is due to a non-functioning zone of the DWCs. 

\section{Assessments of the stochastic term of the energy resolution and of the light confinement }
\subsection{The stochastic term of the energy resolution\label{sec:LYield} }
\begin{figure}[ht!]
    \centering
    \includegraphics[width=0.45\textwidth]{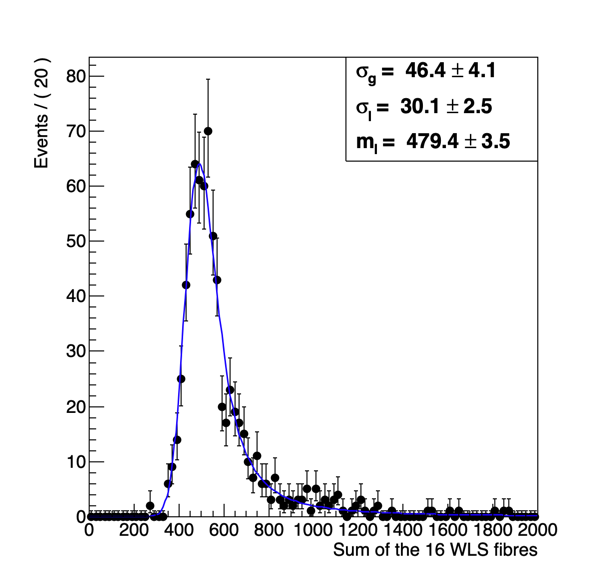}
     \includegraphics[width=0.45\textwidth]{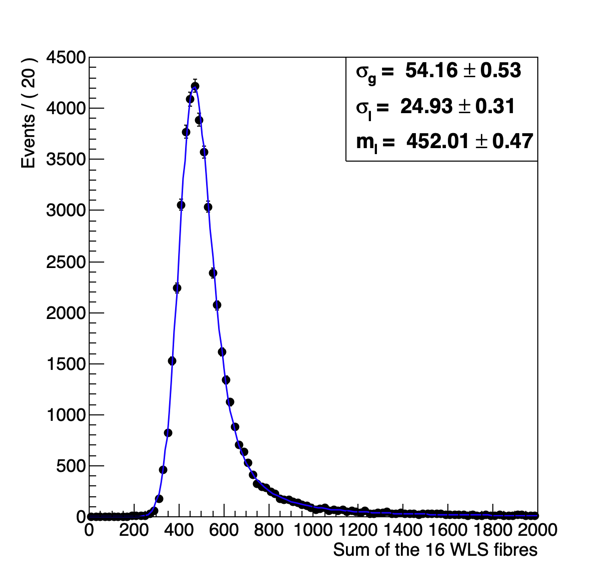}
    \caption{Sum of the number of photo-electrons registered on the 16 fibres for tracks in the four quadrants Q0-3 (see figure~\ref{fig:Homogen}) from fibres 7, 4, 11 and 8. The curve is a Landau distribution convoluted with a Gaussian function. The left plot corresponds to the set-up filled with water and the right plot to the setup filled with the heavy liquid.}
    \label{fig:fitMuonLY}
\end{figure}

An encouraging number of approximately 400 photo-electrons was obtained with the cosmic muons~\cite{Barsuk:2023uva} leading to a stochastic term on the energy resolution of the order of 1\%. The much larger data sample recorded during the test beam and the possibility to better define the tracks used for the measurement calls for a new estimate. In order to avoid border effects, only the events where tracks are hitting the central region of the prototype are used, 
and the data recorded with muon beams is analysed. 
A Landau distribution convoluted with a Gaussian function as described in section~\ref{sec:fitmodel} is fitted to the distribution of the total number of photo-electrons recorded on the 16 WLS fibres. 
The fit for the two periods is shown in figure~\ref{fig:fitMuonLY}.

The number of photoelectrons measured for water and for the heavy liquid are 479 and 452, respectively, and are in very good agreement with the values obtained using cosmic muons.~\cite{Barsuk:2023uva}.

\subsection{Light confinement \label{sec:confL}}
\begin{figure}[ht!]
    \centering
         \includegraphics[width=0.53\textwidth]{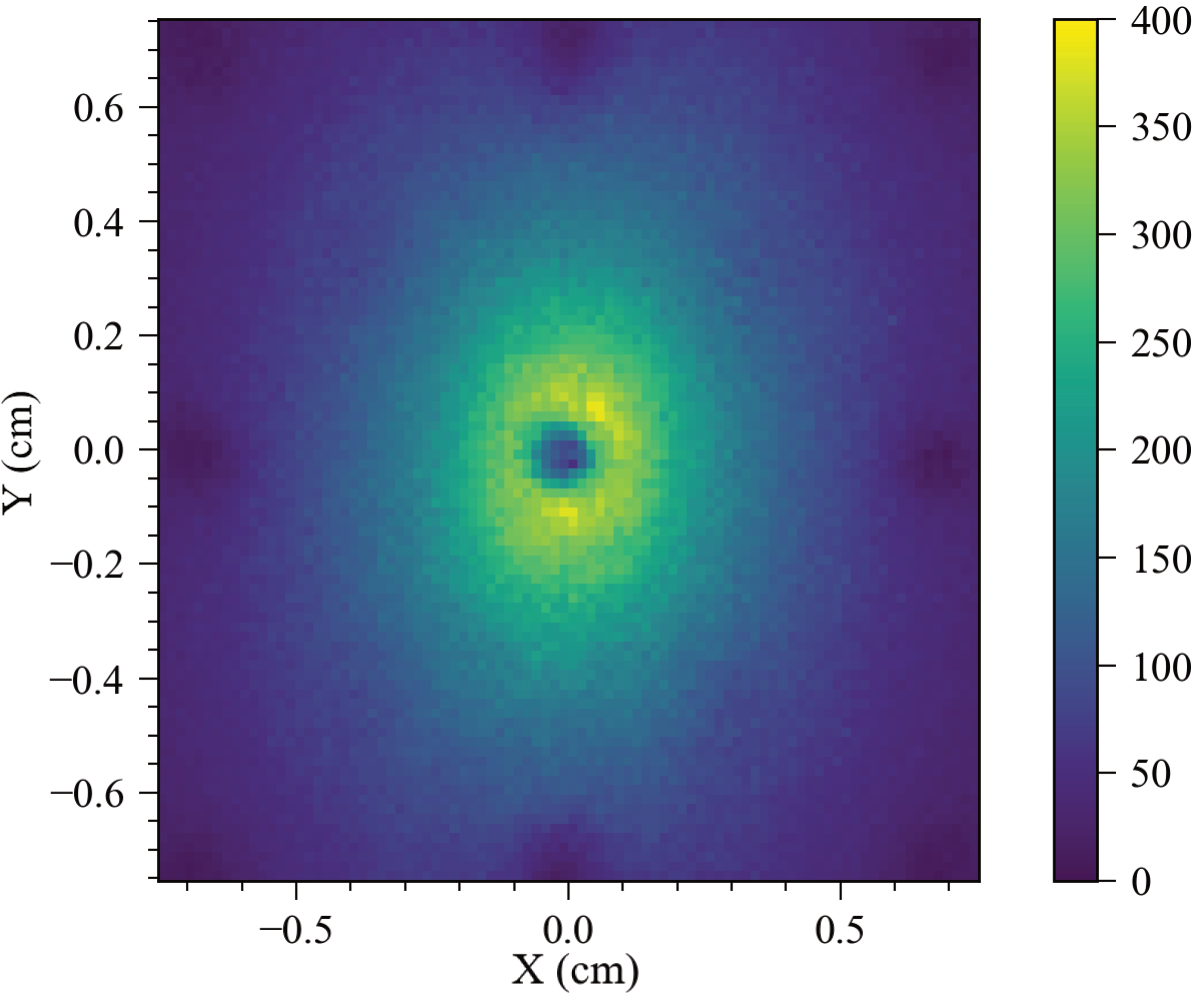}   
    \includegraphics[width=0.43\textwidth]{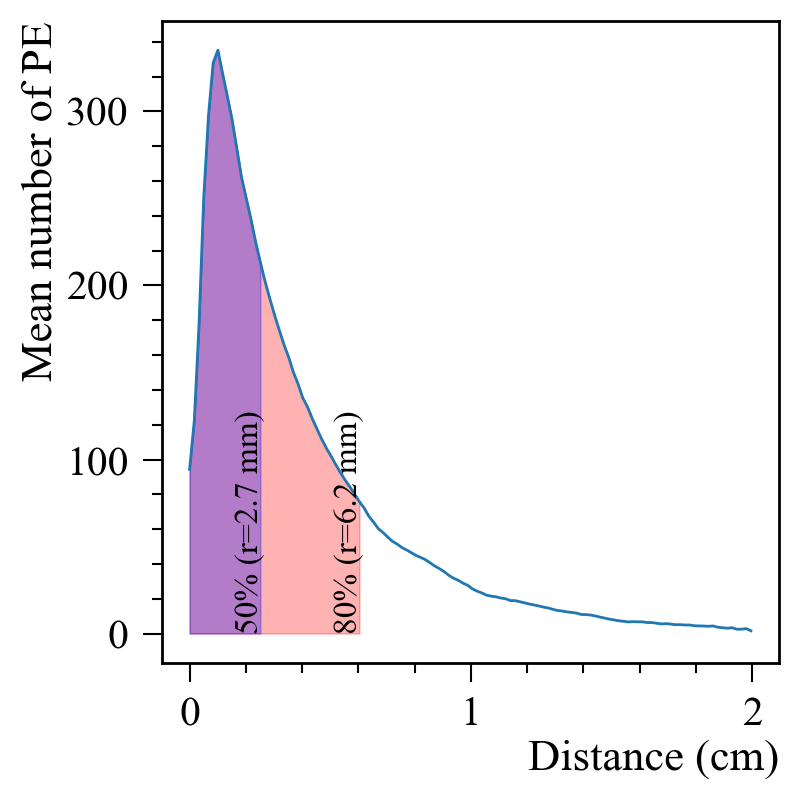}
    \caption{Left: average number of photo-electrons for the four most central fibres. Right: evolution of the mean number of photo-electrons with the distance of the track from the fibre centre.}
    \label{fig:calib:lightC}
\end{figure}

Using the largest statistics run with pions from P2 and the four more central fibres, the average number of photo-electrons read by a given fibre as a function of the position of the track with respect to this fibre is shown in figure~\ref{fig:calib:lightC}. 
These observations clearly confirm that the scintillation light remains confined near its point of production. The right-hand plot displays the dependence of the mean number of photo-electrons on the distance from the fibre centre. Approximately 50\% of the detected light originates from particle interactions occurring within 2.7 mm from the fibre centre, corresponding to slightly less than half the spacing between adjacent fibres. 
These results and those described in section~\ref{sec:hitmap} are used to assess more precisely the fibre position. It is determined via the minimization of the sum of the number of photo-electrons in a square of $3\times3$ bins in a $10\times10$ bin windows around the expected position. The minimum in both directions is taken as the actual fibre position. The precision of this method is about the size of a bin ($170 \ \mu$m).

\section{Analysis framework}

\subsection{Homogenisation of the fibres responses}
\begin{figure}[ht!]
    \centering
     \includegraphics[width=0.45\textwidth]{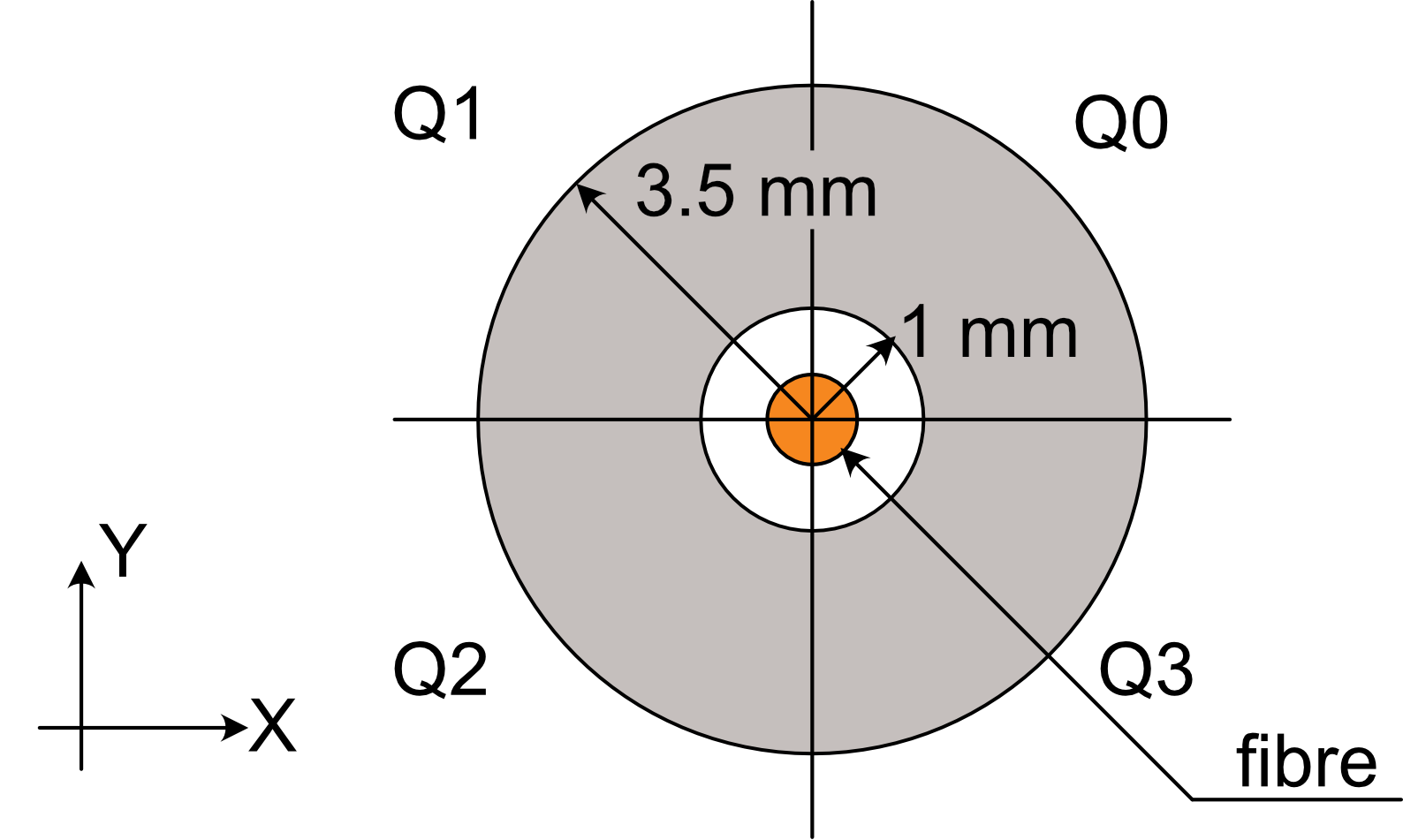}
     \hskip 1cm
    \includegraphics[width=0.35\textwidth]{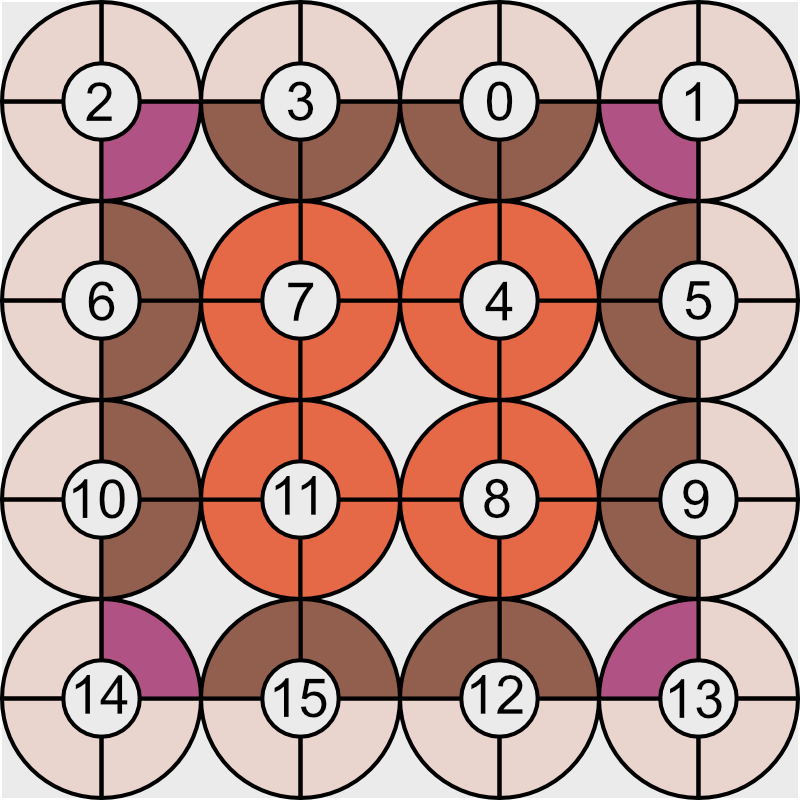}
    \caption{Left: definition of the quarter. The coloured quarters away from the borders of the GRAiNITA prototype are used in the homogenisation procedure.}
    \label{fig:Homogen}
\end{figure}

The analysis is interested in the intrinsic non-uniformity due to possible light absorption as function of the distance between the light emission point and the fibre due to the considerable path length resulting from the reflexion/refraction at the fibre liquid interface. Since during the test beam data-taking period, the fibres have been coupled and decoupled several times to the SiPMs, some additional non-uniformity might therefore have been introduced. In a first step the responses are homogenised using the pions runs from P2. Indeed if the walls are covered by an highly reflective wrapping material, the Vikuiti TM Enhanced Specular Reflector by 3M\footnote{\href{https://www.isoltronic.ch/assets/of-m-vikuiti-esr-app-guide.pdf}{Vikuiti/VM2000 data}} (formerly known as VM2000),
some absorption from the inner walls is present. Consequently, for each fibre, quarters are defined (see figure~\ref{fig:Homogen}-left) and the mean number of photo-electrons for tracks pointing to a given quarter is computed. The arithmetic average of the mean values is taken as a reference, and a coefficient is built in order to reach a homogeneous answer for the whole detector. For each fibre only the quarters that are not adjacent to the detector edges are taken into account in order to avoid boundary effects (figure~\ref{fig:Homogen}-right).

\begin{figure}[tb!]
    \centering
     \includegraphics[width=0.45\textwidth]{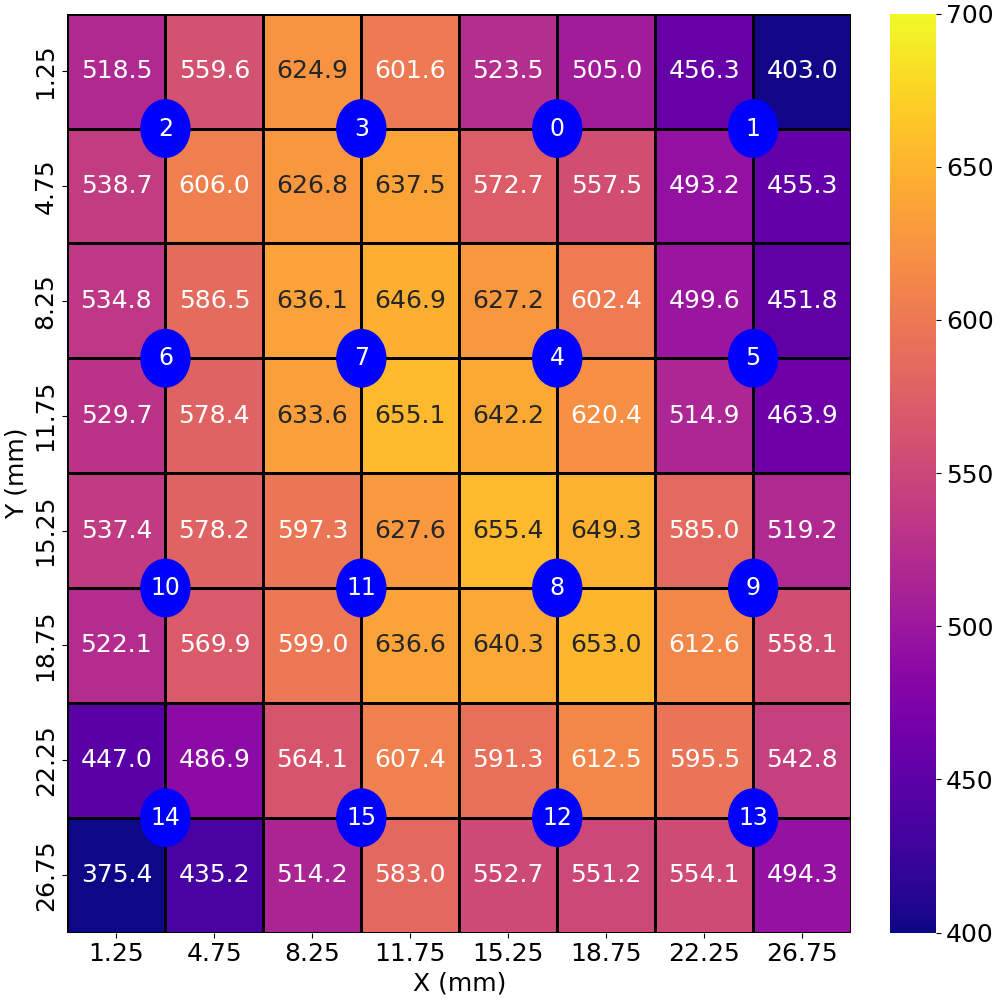}
    \includegraphics[width=0.45\textwidth]{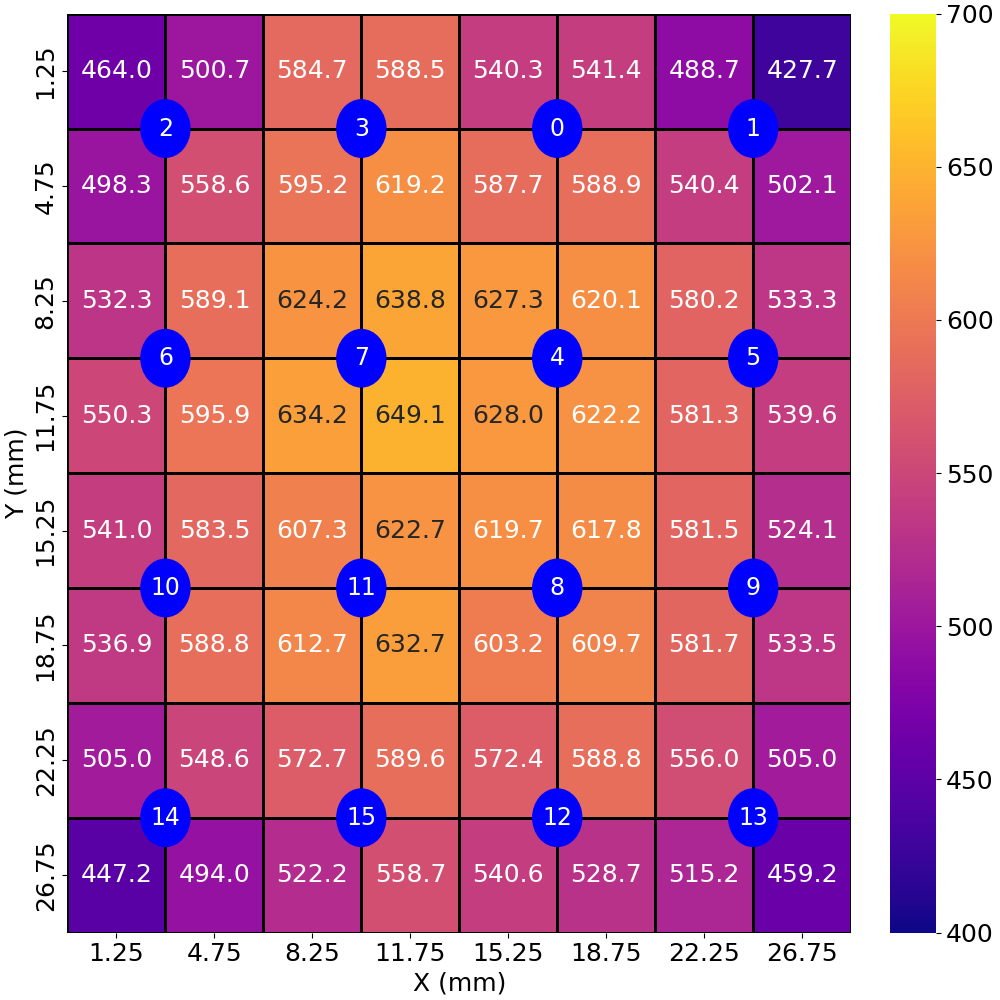}
    \caption{Hitmaps by quarters before (left) and after the homogenisation (right) from the pion runs. The numbers correspond to the mean value of the distribution of the signal obtained for tracks located in each quarter. The blue disks are the locations of the fibres.}
    \label{fig:PionHit_HM}
\end{figure}

In figure~\ref{fig:PionHit_HM} the resulting maps are shown before and after homogenisation. Effects from the vicinity of the prototype walls are visible, already indicating that some fiducial cuts will have to be applied to determine the non-uniformity effect for a scale-one calorimeter. 

\subsection{Design of the pion fitting model \label{sec:fitmodel}}
\begin{figure}[ht!]
    \centering
    \includegraphics[width=0.45\textwidth]{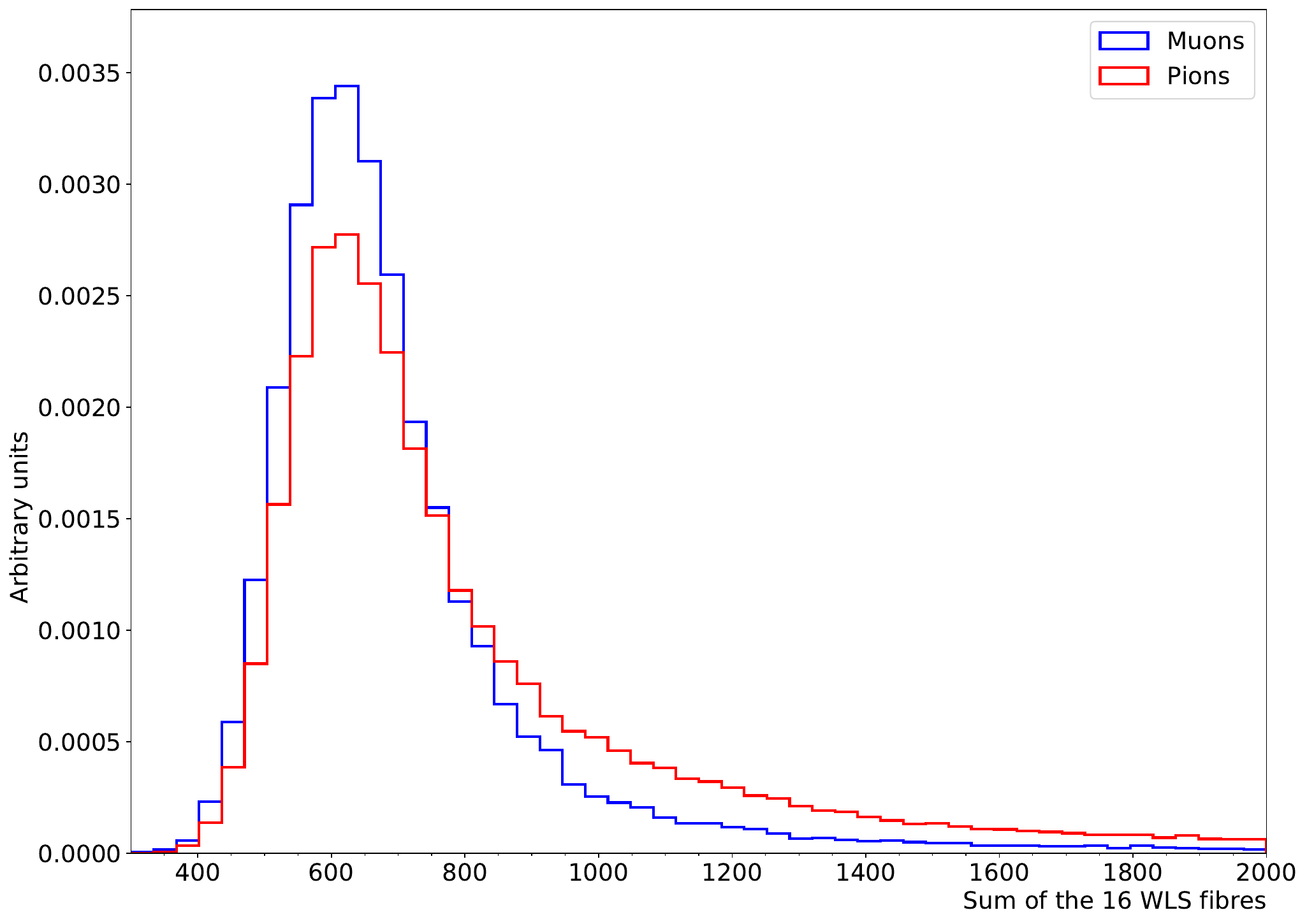}
    \caption{Comparison of the distributions of the total number of photo-electron for tracks located in the quarters nearby the four more central fibres in a muon (blue) run and a pion (red) run.}
    \label{fig:compMuonsPions}
\end{figure}

The measurement of non-uniformity requires a large number of events, and therefore only the pion runs are usable. Consequently, a dedicated model must be developed in order to accurately describe the number of produced photoelectrons. Figure~\ref{fig:compMuonsPions} compares the distributions of the total number of photo-electron summed over the 16 WLS fibres for tracks located in the quarters nearby the four more central fibres in a muon run and a pion run. The higher number of photo-electrons observed in the pion run is attributed to the fraction of pions that undergo interactions within the GRAiNITA prototype.

The runs from P2 are used to design the fitting model for the pion data. As in section~\ref{sec:LYield}, the distribution of the number of photo-electron for the muon data is modelled by the convolution of a  Landau distribution with a Gaussian: 
\begin{equation}
    M_{\mu}(x) = (L(m_l,\sigma_l) \ast G(\sigma_g))(x)
\end{equation}
where $L$ is the Landau distribution with a most probable value $m_l$ and a scale factor $\sigma_l$ and $G$ a Gaussian distribution with a variance $\sigma_g$. The parameter of interest is $m_l$, all the others will be called nuisance parameters in the remaining.

\begin{figure}[ht!]
    \centering
     \includegraphics[width=0.3\textwidth]{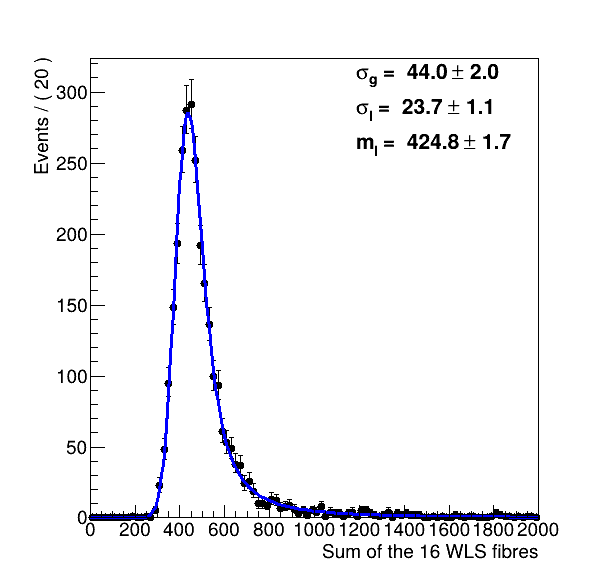}
    \includegraphics[width=0.3\textwidth]{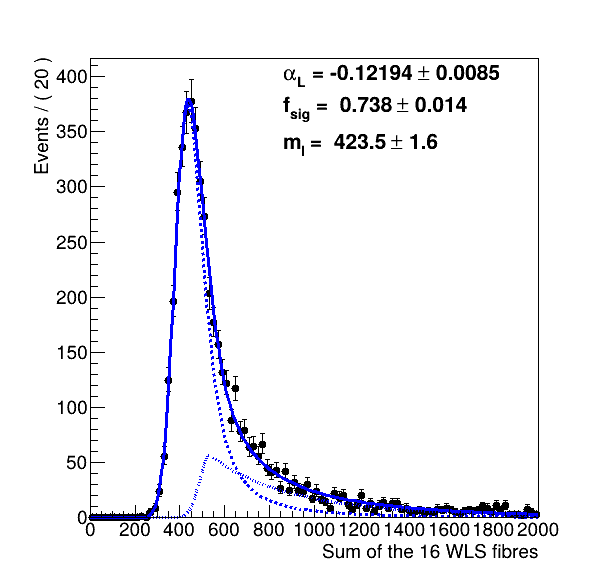}
    \includegraphics[width=0.3\textwidth]{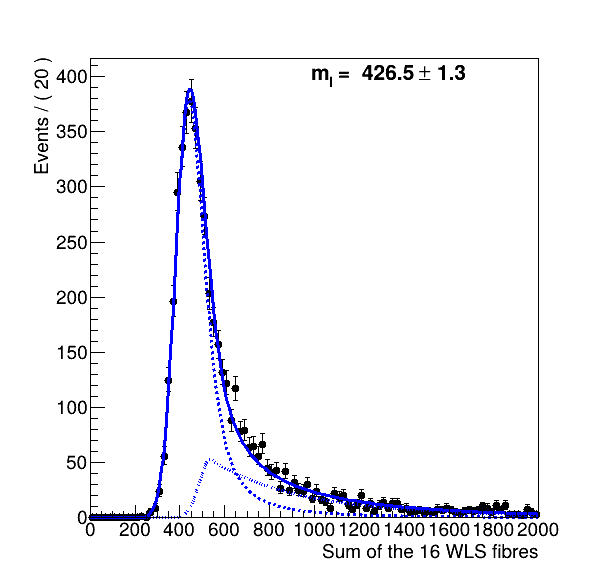}
    \caption{Example of the procedure for tracks in the vicinity of a given fibre. Left: fit of the total number of photoelectrons measured in the muon data from P2. Middle: fit of the total number of photoelectrons measured in pion data from P2. Right: fit for the same pion events after all steps and with $m_l$ as single free parameter.}
    \label{fig:compPoModel}
\end{figure}

In order to describe the pion distribution, a Crystal Ball distribution $CB$ is added:
\begin{equation}
    M_{\pi}(x) = f_{\mathrm{sig}}M_{\mu}(x) + (1-f_{\mathrm{sig}}) CB(\sigma_L, \alpha_L, n_L, x_0)(x) 
\end{equation}
where $\alpha_L$, $\sigma_L$ and $n_L$ are the Crystal Ball shape parameters and $x_0$ the peak position. Some of the Crystal Ball parameters are specifically set to $x_0 = 8/9\sqrt{2}m_l$, $\sigma_L= \sigma_g$ and $n_L = 6$. The values $\sigma_l$ and $\sigma_g$ are obtained from the muon fit. The free parameters are then $m_l$, $\alpha_L$ and $f_{\mathrm{sig}}$. The principle of the method is shown in figure~\ref{fig:compPoModel}.

In order to determine a global and unique set of mean values for all the nuisance parameters which can be used for the fibres, fits for tracks in the vicinity of the 4 more central fibres were performed and the distributions of the values of the nuisance parameters were constructed for well-behaved fits. In the following the nuisance parameters values are fixed to the  mean values of these distributions given in table~\ref{tab:nuisanceP}. The only free parameter is then the most probable value of the Landau distribution, $m_l$.

\begin{table}
    \centering
    \begin{tabular}{c|c|c|c}
   $\sigma_l$      & $\sigma_g$  & $f_{\mathrm{sig}}$  & $\alpha_L$\\ \hline
19.7      & 49.6 & 0.736 & -0.126 

    \end{tabular}
    \caption{Values of the nuisance parameters of the fit.}
    \label{tab:nuisanceP}
\end{table}
The three steps of the model building and the result are shown in figure~\ref{fig:compPoModel} with the final result on the right-hand plot. A very good description of the distribution is obtained. The values of $m_l$ obtained for muon and pion tracks in the four quadrants around the 16 fibres are compared. They are found to be in agreement with a relative uncertainty of 1.2 \% that is consequently assigned as the method systematic uncertainty, commensurate to its statistical uncertainty.  

\section{Impact of the detector non-uniformity\label{sec:nonHumiformity}}
While the homogenization and fitting model design steps were implemented using P2 data, the non-uniformity measurement is conducted using P1 data owing to its higher quality. Indeed, the non-uniformity behaviour of GRAiNITA is expected not to depend significantly from the liquid (water or water-based sodium polytungstate) nature given the similarity of their refractive indices (respectively 1.33 and 1.5).   
\subsection{Design of the virtual units}
\begin{figure}[tb!]
    \centering
    \includegraphics[width=0.7\textwidth]{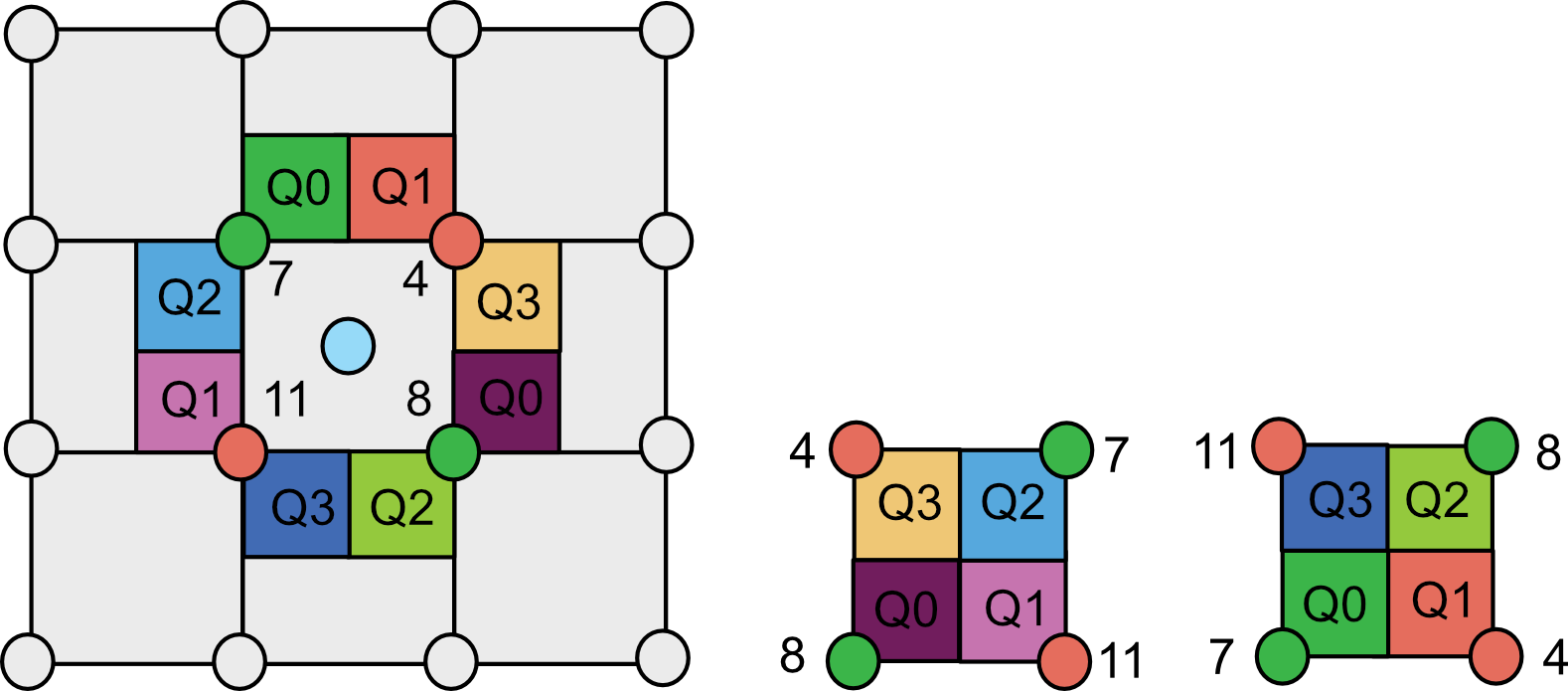}  
    \caption{Left: selected quarters for the virtual unit. Right: the two virtual units which are built from these quarters.}
    \label{fig:virtualUnits}
\end{figure}

 Given the reduced size of the prototype, the regions close to the walls can not be used nor the regions close to the central clear fibre. To circumvent this issue, the information from tracks pointing to different regions is combined into two "virtual units" as illustrated in figure~\ref{fig:virtualUnits}.

\subsection{Determination of the uniformity maps}
\begin{figure}
    \centering
    \begin{overpic}[width=0.32\linewidth]{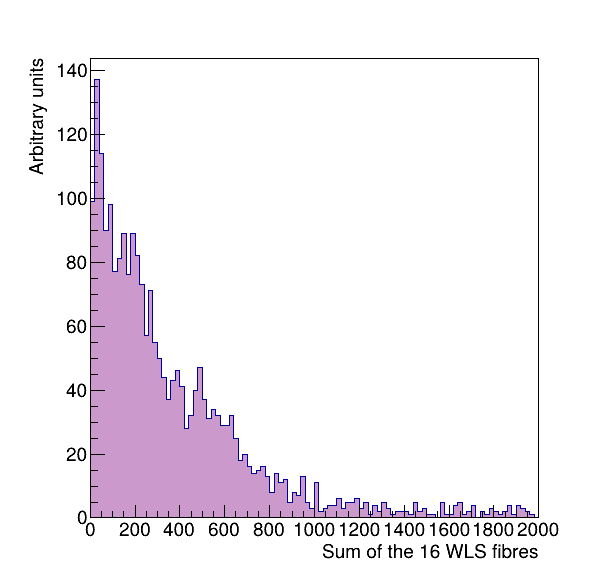}
        \put(50,50){\includegraphics[width=0.1\linewidth]{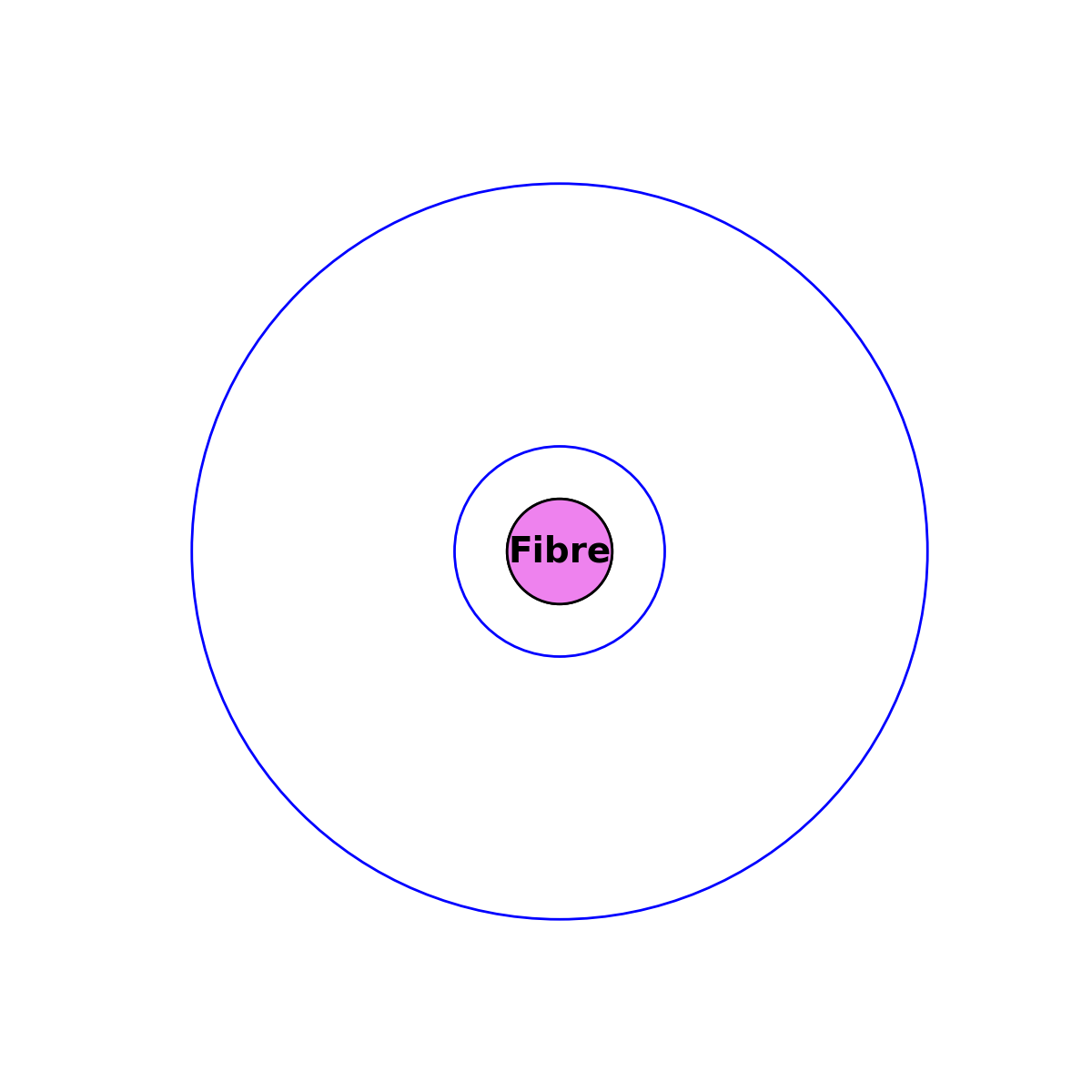}}
    \end{overpic}
    \begin{overpic}[width=0.32\linewidth]{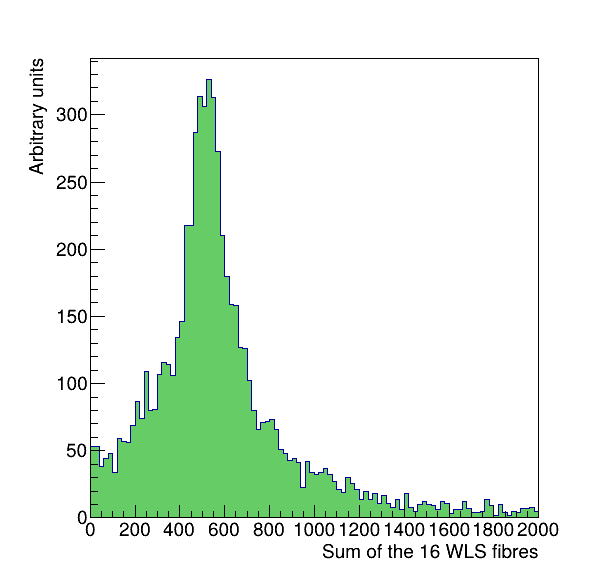}
        \put(50,50){\includegraphics[width=0.1\linewidth]{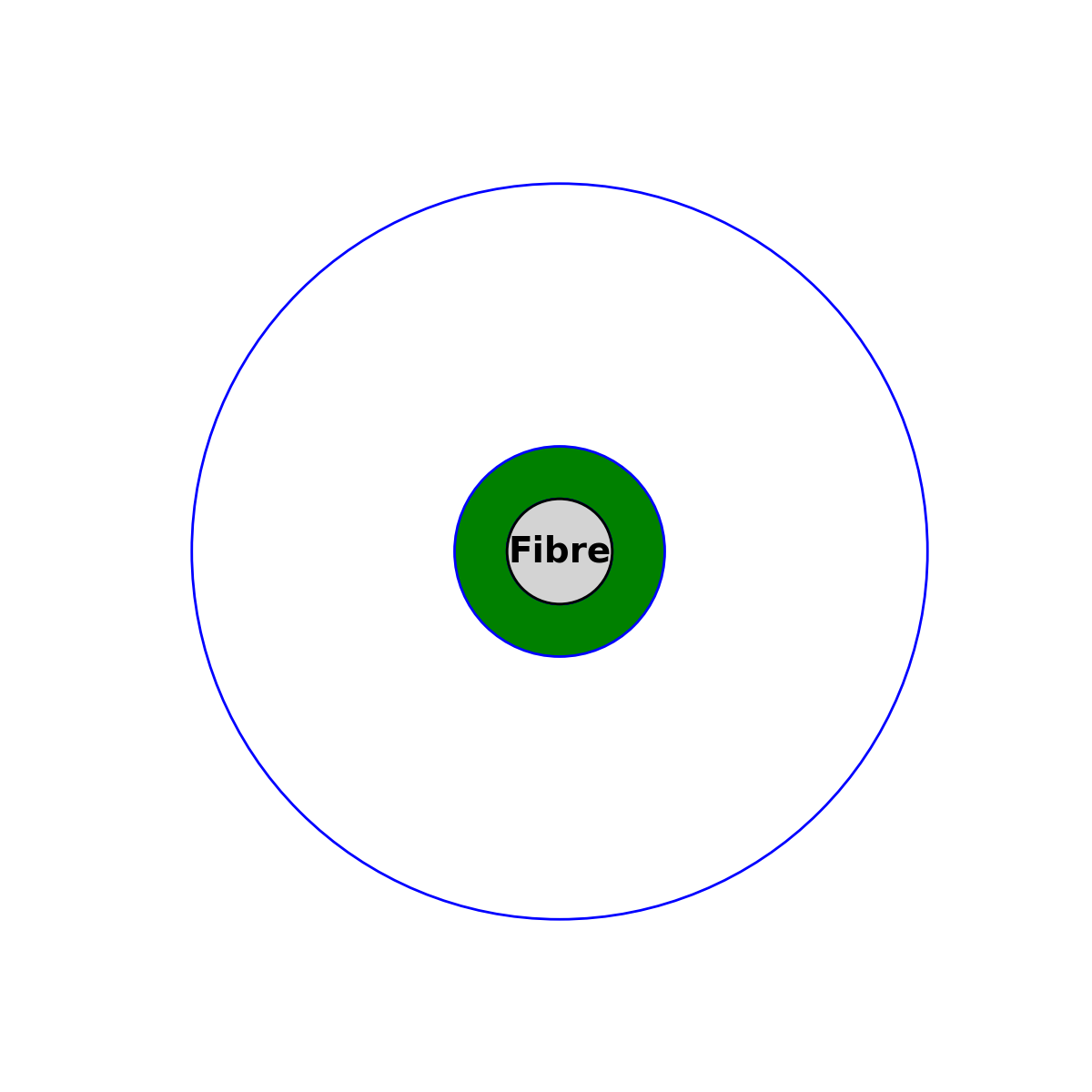}}
    \end{overpic}
    \begin{overpic}[width=0.32\linewidth]{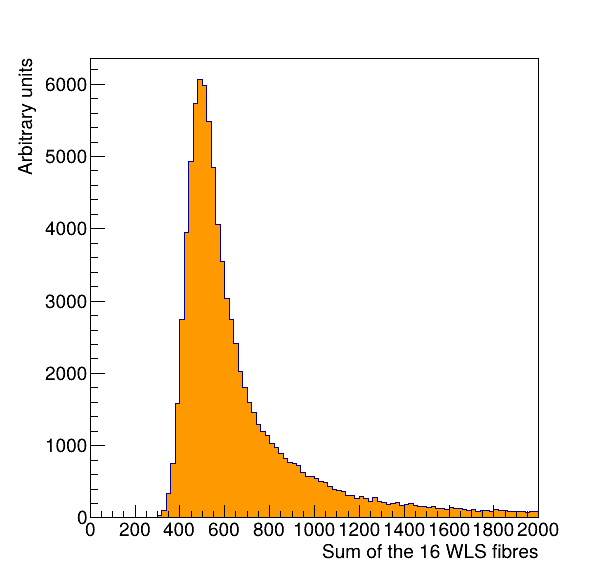}
        \put(50,50){\includegraphics[width=0.1\linewidth]{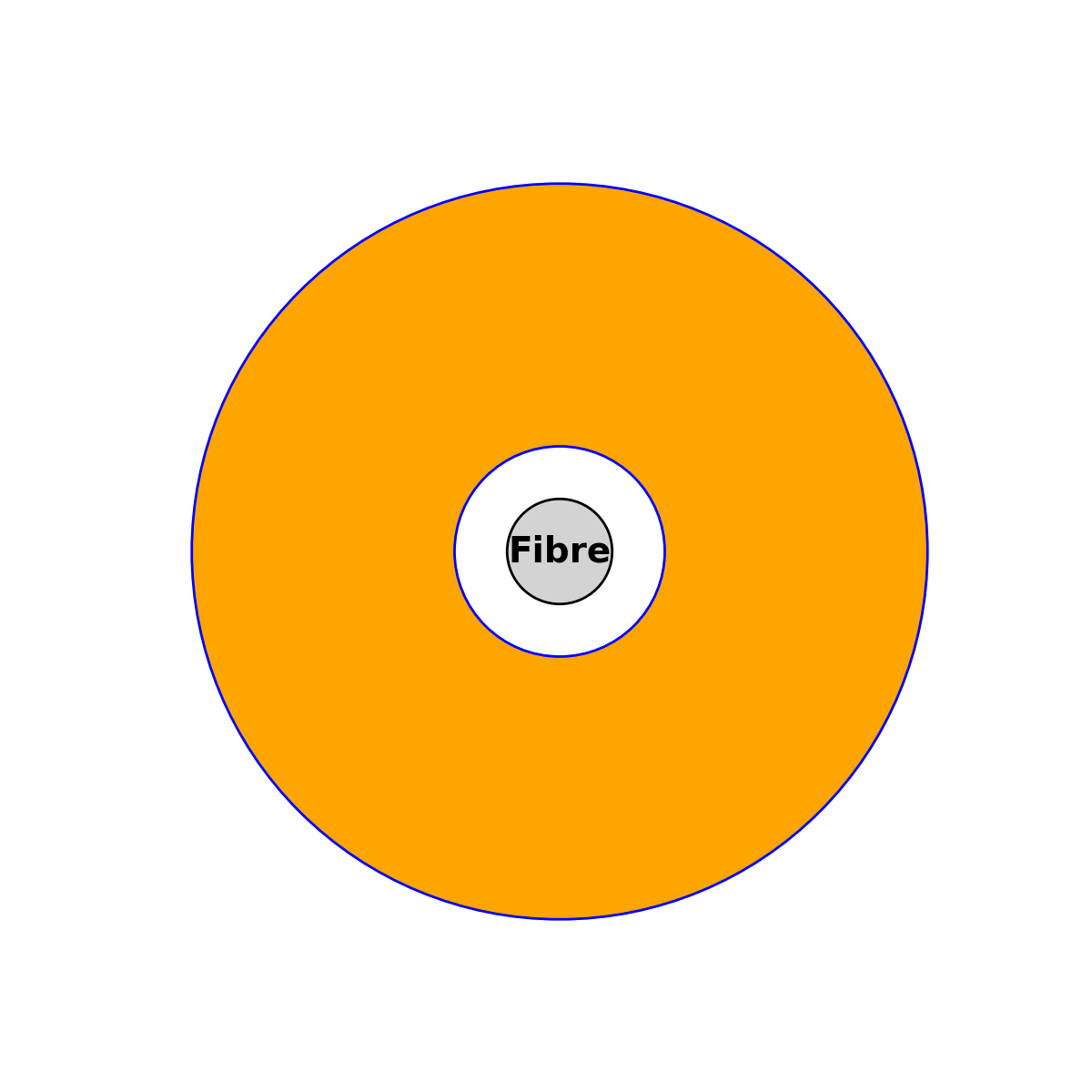}}
    \end{overpic}
     \caption{From left to right: signal recorded by the sum of the 16 WLS fibres when tracks are selected in regions centred on a fibre (violet disk), very close to a fibre (green corona, annulus of inner radius 0.5 mm, outer radius 1 mm) or further apart from a fibre (orange corona, annulus of inner radius 1 mm, outer radius 3.5 mm).}
    \label{fig:TrackInOutFiber}
\end{figure}

A two-dimensional scan with steps of 
1 mm$^2$ is performed. It should be noted that when tracks fall within a bin containing a fibre, the resulting signal differs significantly (see figure~\ref{fig:TrackInOutFiber}). To mitigate this effect, the values of the bins containing a WLS fibre are replaced with the mean of the adjacent bin values. The absence of scintillation light for tracks pointing to a WLS fibre will be taken into account in the simulation described in section~\ref{sec:simu-nonuniformity}.

\begin{figure}[htb!]
    \centering
    \includegraphics[width=0.45\textwidth]{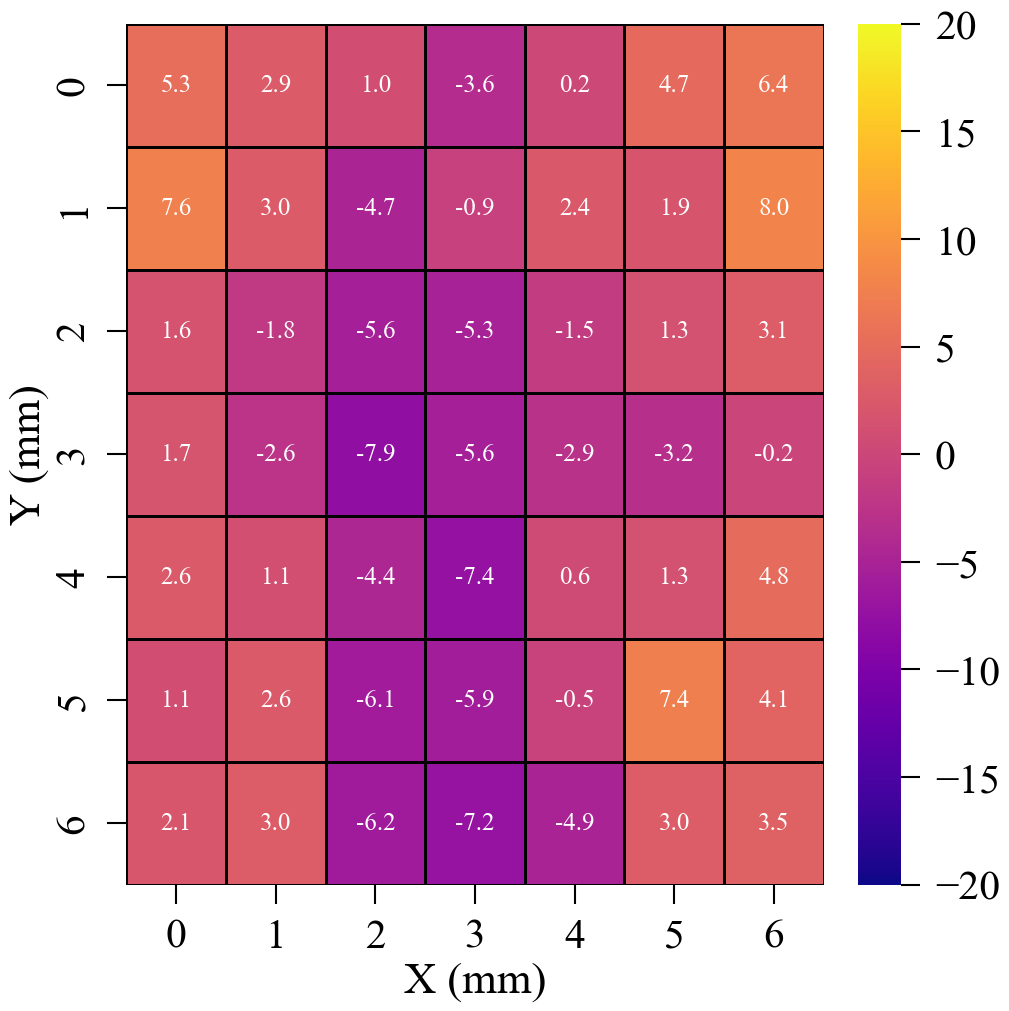}
     \includegraphics[width=0.45\textwidth]{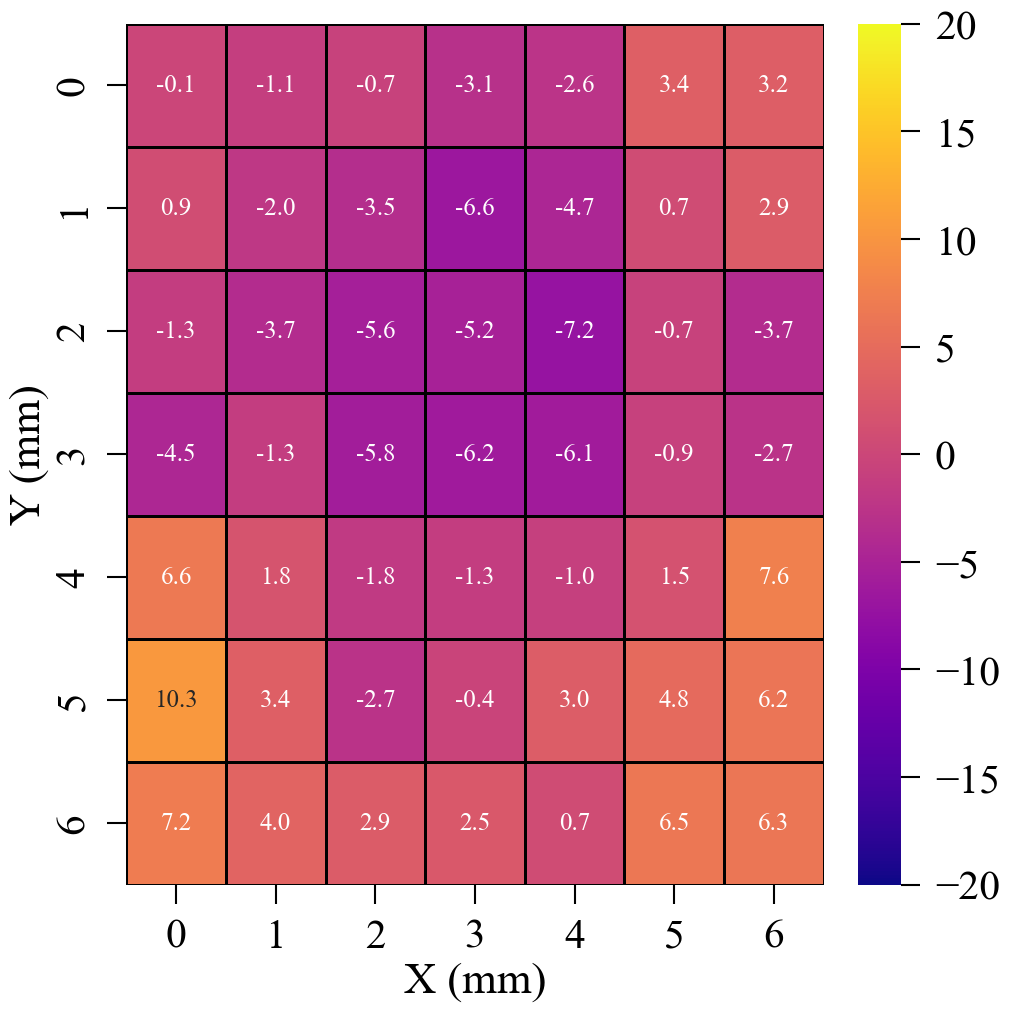}
    \caption{Uniformity maps for the two virtual units. The values are indicated in percent relative to the average.}
    \label{fig:virtualRelative}
\end{figure}

The resulting two uniformity maps expressed as the values relative to the mean are shown in figure~\ref{fig:virtualRelative}.

\subsection{Impact of the non-uniformity response on the constant term \label{sec:simu-nonuniformity}}
A simplified model of a full scale calorimeter module is simulated using Geant4 Monte-Carlo~\cite{G4}. The size of the calorimeter module is $168 \times 168 \times 400$~mm$^3$. The depth of 400 mm corresponds to 25 radiation lengths ($X_0$), which is sufficient to fully contain electromagnetic showers initiated by 25 GeV photons. This volume contains a grid of 24 by 24 WLS fibres evenly distributed in the corners of squares of 7 by 7 mm$^2$, to reproduce the small-scale GRAiNITA prototype. The WLS fibres are simulated as cylinders of 1 mm diameter. In order to assess in a reasonable amount of CPU time the impact of the measured non-uniformity the grains are not simulated individually but a single material was simulated with partial contents of the scintillator and heavy liquid corresponding to the ones measured with the physical detector used for the beam test. The mass fractions of the detector material simulated are : 20.77 \% of heavy liquid and 79.23 \% of ZnWO$_4$. The radiation length, calculated with the Geant4 framework, of this mixture of materials is 1.64 cm which aligns with manual cross-calculations.  

\begin{figure}[htbp]
    \centering
    \includegraphics[width=1\linewidth]{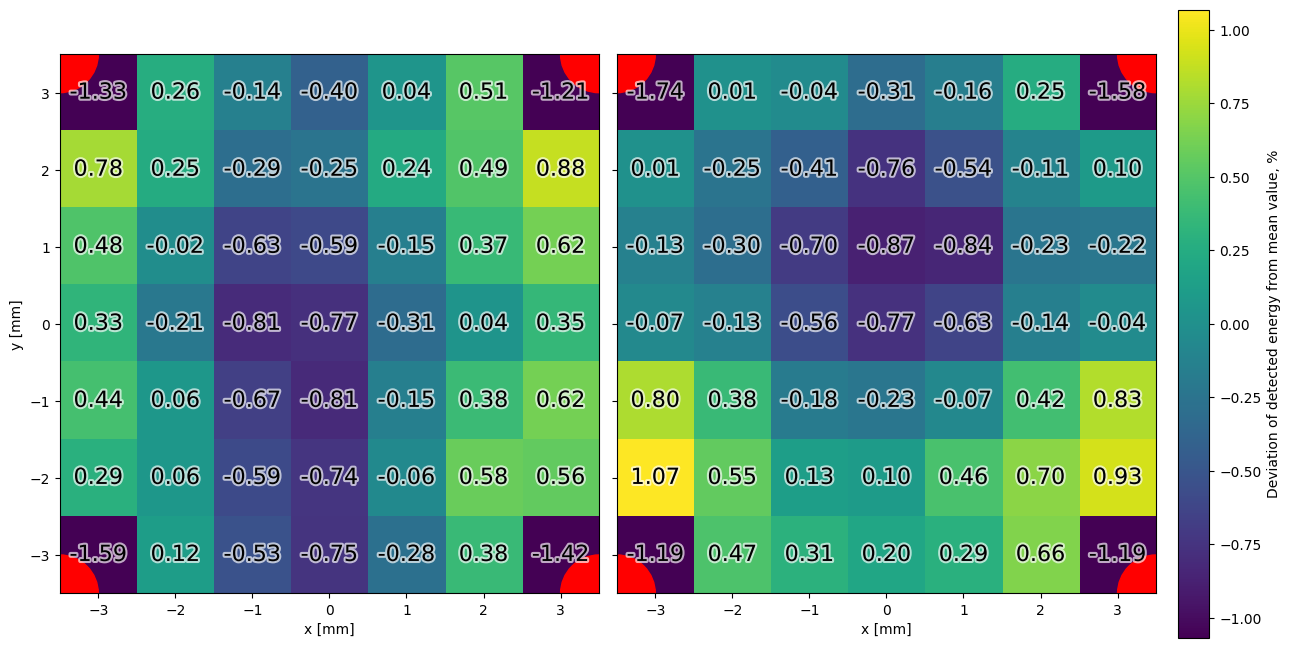}
    \caption{The maps of the position-dependent deviation of the effective detected energy from the mean value for a 25 GeV incoming photon. The locations of the WLS fibers are noted with red color.}\label{fig:G4maps}
\end{figure}

A 25 GeV incoming photon with a position uniformly distributed on the calorimeter surface is generated. The detector is virtually split into rectangular volumes of $1 \times 1 \times 400$~mm$^3$. Because WLS fibre material does not scintillate, the energy deposited in WLS fibre is not taken into account for the corresponding strip. The effective energy is estimated from the deposited energy in a strip multiplied by the efficiency from this strip. More specifically, the efficiency comes from the maps 
shown in figure~\ref{fig:virtualRelative}. 
This simplified simulation assumes that the light detection efficiency is independent of the position along the fibres.
To estimate the contribution to the energy resolution, the event is simulated with primary hit positions evenly distributed on the surface between 4 fibres at the centre of the detector. The effective detected energy is calculated for each simulated event. The procedure is performed for the 
two experimental maps obtained from the test beam (see figure~\ref{fig:virtualRelative}).  
The resulting maps of the position-dependent deviation of the effective detected energy from the mean value are shown in figure~\ref{fig:G4maps}. 

\begin{figure}[htbp]
\centering
\includegraphics[width=0.45\textwidth]{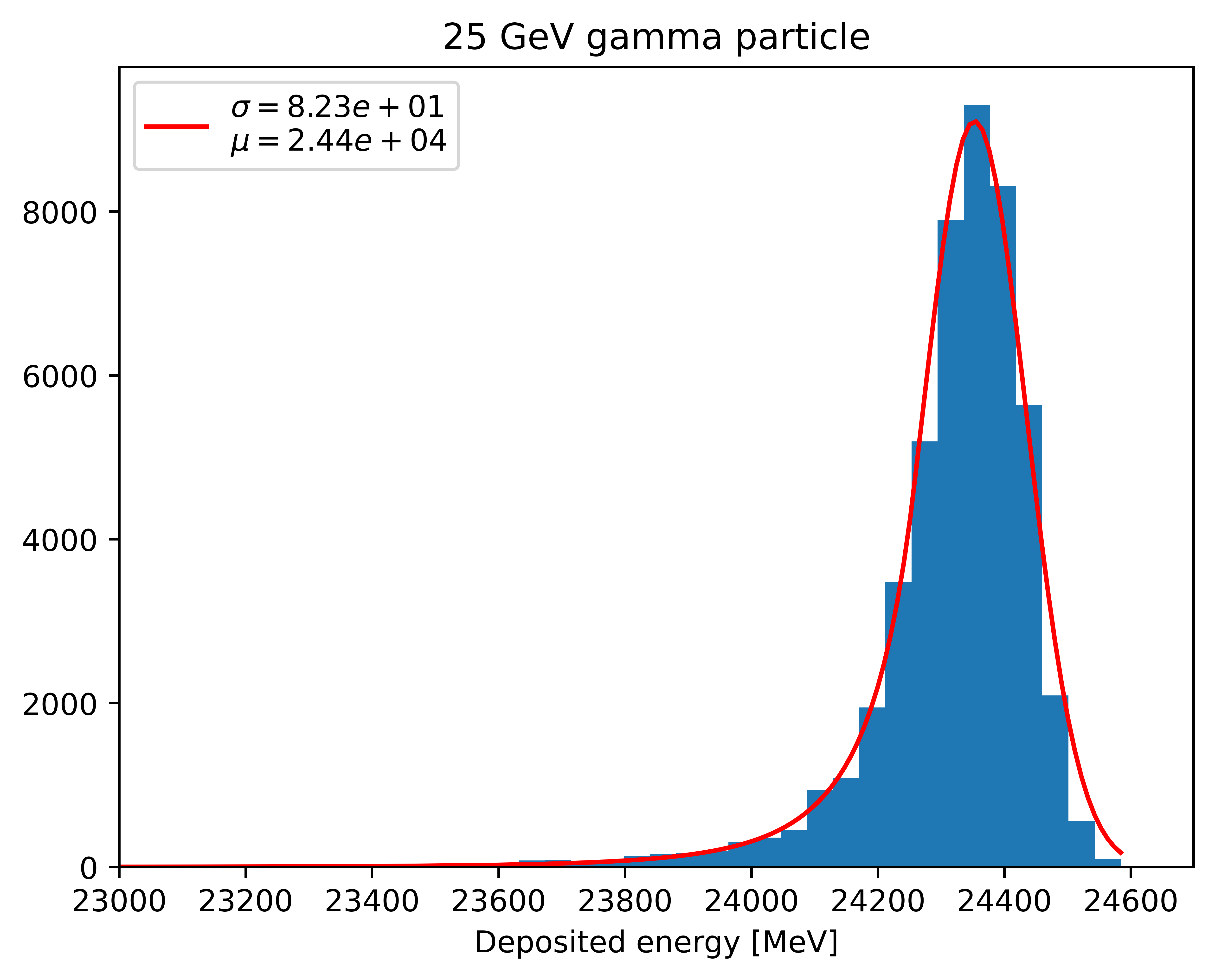}
\includegraphics[width=0.45\textwidth]{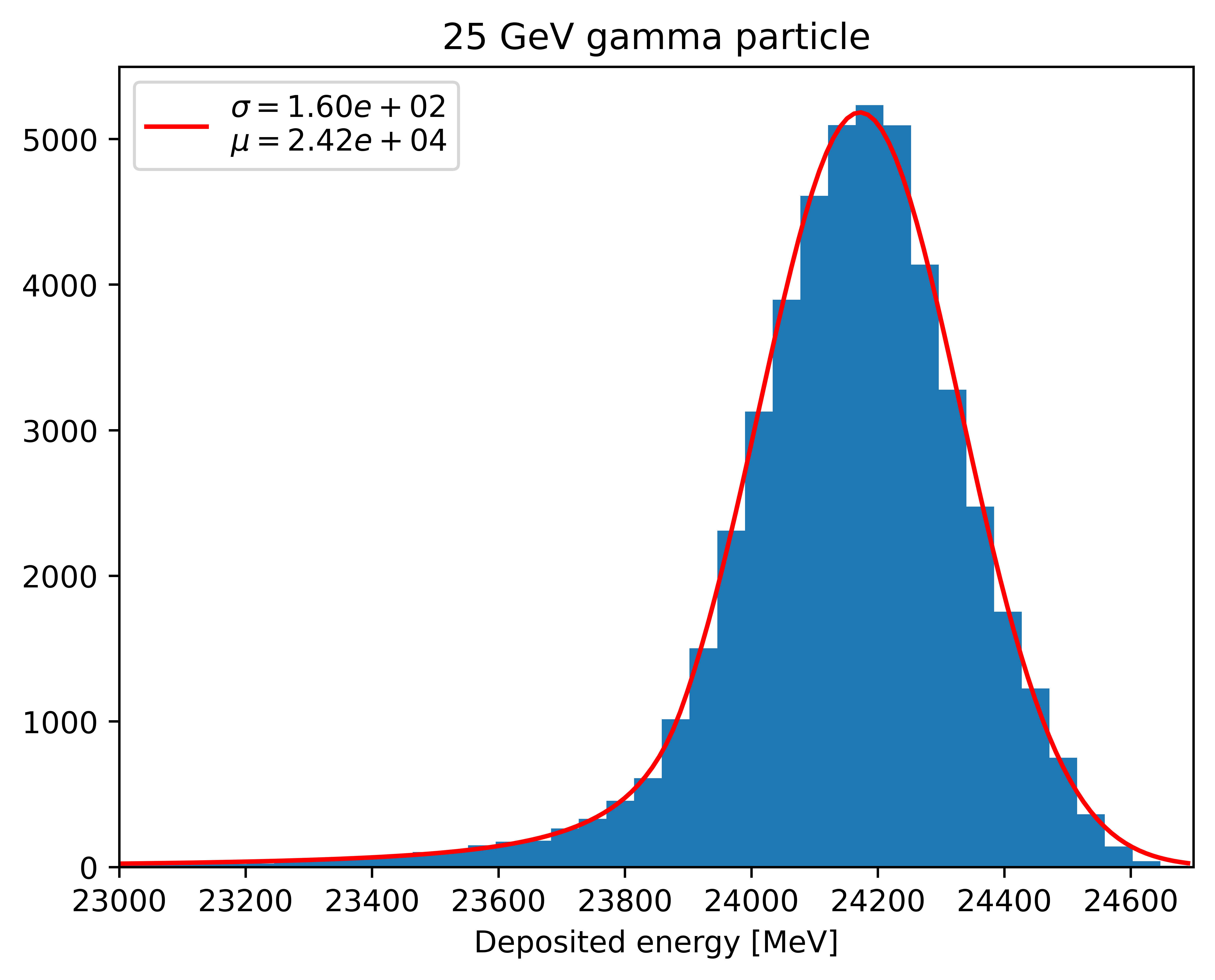}
\caption{Deposited energy of a 25 GeV incoming photon for a fully uniform module without WLS fibres (left) and a module including WLS and the maps extracted from the test beam data (right).}
\label{fig:G4DepositedE}
\end{figure}

In order to estimate the overall impact on the energy resolution, the following procedure is used. The energy deposited for a module without WLS fibres and accordingly a fully uniform map is first obtained (figure~\ref{fig:G4DepositedE}-left). The width of the distribution is due to the fluctuation of the particles escaping the calorimeter module.
A similar distribution is built from the realistic module with WLS fibres and the maps shown in figure~\ref{fig:virtualRelative} is computed (figure~\ref{fig:G4DepositedE}-right). The comparison of the widths of the two distributions allows to extract a constant term of the order of 0.7\% (including the systematics uncertainty estimated from the pion model in section ~\ref{sec:fitmodel} due to the non-uniformity of the module). It was checked that this constant term does not depend on the energy.

\section{Conclusion}
Data from a two-day period in June 2024 using the H9 test beam of the SPS at CERN recorded with the small-scale GRAiNITA prototype is analyzed. These measurements confirm that the contribution to the energy resolution due to the number of photo-electrons is of the order of $1\%/\sqrt{E}$. With preliminary simulations of the grains and the heavy liquid~\cite{bib:IEEE2021}, this term goes to about $2\%/\sqrt{E}$. More importantly, the data recorded during this test beam period enables a first estimation of the constant term associated with detector non-uniformity. Despite the limitations of these measurements due to the small size of the detector and the use of pions, the constant term due to the non-uniformity should be significantly below 1\%.
\acknowledgments
We would like to thank Kuraray for providing WLS fibre samples, David Simpson from Pangea Ltd for useful information about the water-based sodium polytungstate solution, and Loris Martinazzoli (CERN) and his team for having welcome us during the PicoCal test beam. Finally we'd like to gratefully thank Yuri Guz who patiently accompanied our experimental efforts and provided a reliable trigger and tracking system in front of the prototype.


\bibliographystyle{JHEP}
\bibliography{main.bib}

\end{document}